\documentstyle[preprint,12pt,aps]{revtex} 
\title{The Atomic Limit of the Boson Fermion Model}
\author{T.~ Domanski$^{\dagger}$, J.~Ranninger and J. M. Robin}

\address{Centre de
Recherches sur les Tr\`es Basses Temp\'eratures, Laboratoire
Associ\'e \'a l'Universit\'e Joseph Fourier, Centre National de la
Recherche Scientifique,\\ BP 166, 38042, Grenoble C\'edex 9, France}

\date{\today} 
\begin{document} 
\maketitle 
\draft 
\begin{abstract}
The Boson-Fermion model, describing a mixture of hybridized localized 
Bosons and itinerant Fermions on a lattice, is known 
to exhibit spectral properties 
for the Fermions which upon lowering the temperature develop into 
a three pole structure in the 
vicinity of the Fermi level. These spectral features go hand in hand with 
the opening of a pseudogap in the density of states upon approaching 
the critical temperature $T_c$ when 
superconductivity sets in. In the present work we study this model, in
the atomic limit where the three pole structure arises naturally from the 
local bonding, anti-bonding and non-bonding states between the Bosons 
and Fermions.


\end{abstract} 

\newpage 

The Boson Fermion model presents one of the possible scenarios  which 
are presently 
widely discussed in connection with the phenomenon of High $T_c$ 
superconductivity\cite{exp}. 
This model treats localized Bosons (such as bi-polarons)
which are locally
hybridized with pairs of itinerant Fermions (band electrons). 
On the basis of this model one envisages the 
opening of a pseudogap in the density of states  of the electrons 
as the temperature decends below a certain characteristic temperature 
$T^*$\cite{Batlogg-94} and approaches $T_c$ from above. 
The opening of the pseudogap has been 
attributed to a precursor effect of superfluid phase 
fluctuations\cite{spectral} with the amplitude of the order parameter 
remaining finite. Near the Fermi level this 
is tantamount to the existence of resonance two-particle states into which 
most of the electrons are engaged. The physical implications of these 
one-particle features on the thermodynamic, transport and magnetic properties 
demonstrate this behaviour, which
radically differs from that expected for classical Fermi liquids\cite{thermo}.
The deviations from Fermi liquid properties are manifest in a complex
three-pole structure of the one-particle electron Green's 
function\cite{spectral}. 
These findings have been obtained by solving selfconsistently the equations 
for the Boson and Fermion selfenergies using a diagramatic conserving 
approximation, a procedure from which the underlying physics could be infered. 

In this paper, we show that some of these physical interpretations can 
indeed be put on a firm basis when studying  the atomic limit of this model.
In this case, the local interaction between a hard core Boson
and a pair of Fermion (having opposite spins) on each site of the
lattice can be diagonalized exactly. The kinetic part of the Fermions
is subsequently treated in an approximation \`a la Hubbard.

The Boson-Fermion hamiltonian in the Wannier representation is given
by\cite{Physica}
\[
H \; = \; (zt - \mu) \; \sum_{j,\sigma} \; c_{j,\sigma}^{\dagger}
  c_{j,\sigma}
  \; - t \; \sum_{j,\delta,\sigma} \; c_{j+\delta,\sigma}^{\dagger}
  c_{j,\sigma}
  \; + \; (\Delta_{_B} - 2 \mu) \; \sum_{j} \; b_{j}^{\dagger} b_{j}
\]
\begin{equation}
\; \; \; \; \; \; + \;
v \; \sum_{j} \; \left[ \; c_{j,\uparrow}^{\dagger} c_{j,\downarrow}^{\dagger}
b_{j} \; + \; b_{j}^{\dagger} c_{j,\downarrow} c_{j,\uparrow} \; \right],
\end{equation}
where the sum over $\delta$ runs over the $z$ nearest neighbours.
$\mu$ denotes the chemical potential, $\Delta_{_B}$  the localized Boson level, 
$t$  the hopping integral for the Fermions and $v$ the strength of the 
local hybridization.
The Fermionic operators $c_{j,\sigma}^{(\dagger)}$ anticommute while the 
bosonic operators $b_{j}^{(\dagger)}$ commute for different sites and 
anticommute for the same site.
The Hilbert space for each site is then the direct product of the
Fermionic and Bosonic states:
\begin{equation}
\left\{ \; |\; 0 > , \; |\; \uparrow > , \; |\; \downarrow >, \; |\;
\uparrow \downarrow > \; \right\} \; \otimes \; \left\{ \; |\; 0 ) , \;
|\; 1 ) \; \right\}.
\end{equation}
The local interaction term mixes the states
$|\;\uparrow\downarrow>\otimes|\;0)$ and
$|\;0>\otimes|\;1)$. The corresponding bonding and anti-bonding eigenstates are
\begin{equation}
\left\{ \begin{array}{lclcl}
| \; B^{-} > & = & U \; | \; \uparrow \downarrow > \otimes | \; 0 )
	     & - & V \; | \; 0 \; > \otimes | \; 1 )
\\ \\
| \; B^{+} > & = & V \; | \; \uparrow \downarrow > \otimes | \; 0 )
	     & + & U \; | \; 0 \; > \otimes | \; 1 )
\end{array}
\right.
\end{equation}
with the associated eigenvalues $\varepsilon_{\pm} = 
(2\varepsilon_0 +E_0)/2 \pm\gamma$.
\begin{equation}
U^{2} \; = \; \frac{1}{2}
	   \left( 1 - \frac{2 \varepsilon_{0} - E_{0}}{2 \; \gamma} \right)
\; \; \; \mbox{,} \; \; \;
V^{2} \; = \; \frac{1}{2}
	   \left( 1 + \frac{2 \varepsilon_{0} - E_{0}}{2 \; \gamma} \right)
\; \; \; \mbox{,} \;\;\; UV \; = \; \frac{v}{2 \; \gamma}
\end{equation}
and $\varepsilon_{0} = zt - \mu$, $E_{0} = \Delta_{_B} - 2\mu$ and
$\gamma = \sqrt{(\varepsilon_{0} - E_{0}/2)^{2} + v^{2}}$.
In the atomic limit the one-particle Fermionic local Green's
function
\(
G_{Loc}(\tau) = \; - < T \; [ \; c_{j,\uparrow}(\tau) \;
c_{j,\uparrow}^{\dagger} \; ] >
\)
is then given by
\begin{equation}
G_{Loc}(i\omega_n) \; = \; \frac{Z_{f}}{i\omega_n - \varepsilon_{0}} \; + \;
\frac{1 - Z_{f}}{i\omega_n - v^{2}\; / \; 
( i\omega_n + \varepsilon_{0} - E_{0} ) }
\end{equation}
where $Z_{f}$ is the temperature dependent factor 
\begin{equation}
Z_{f} \; = \; \frac{1 \; + \; e^{-\beta \varepsilon_{0}}  \; \; 
         + \: e^{-\beta(\varepsilon_{0} + E_{0})}
      \; + \; e^{-\beta(2 \varepsilon_{0} + E_{0})}}{
		  1 + 2 e^{-\beta \varepsilon_{0}}
		    + 2 e^{-\beta(\varepsilon_{0} + E_{0})}
		    + e^{-\beta(2 \varepsilon_{0} + E_{0})}
		    + e^{-\beta(\varepsilon_{0} + E_{0}/2 - \gamma)}
		    + e^{-\beta(\varepsilon_{0} + E_{0}/2 + \gamma)}}
\end{equation}
and $\beta = 1/{k_B T}$ denotes the inverse of the temperature.
This Green's function contains three poles with  spectral weights $Z_{f}$, 
$U^{2}(1-Z_{f})$ and $V^{2}(1-Z_{f})$. 
The first term in the expression for this Green's function, Equ.(5), 
determines the contribution from the non-bonding 
one-particle Fermionic states, while the second term is reminiscent of 
Bogoliubov type excitations and contains the effect of hybridized 
two-particle bonding and respectively anti-bonding states, 
$| \; B^{-} >$ and $| \; B^{+} >$.
In the remainder of this paper we consider the symmetric case i.e., 
$\varepsilon_{0} = 0$ 
and $E_{0} = 0$ for which
\(
< c_{j,\sigma}^{\dagger} c_{j,\sigma} > \; = \;
< b_{j}^{\dagger} b_{j} > \; = \; 1/2.
\)
We then obtain
$Z_{f} = 2/(3 + \cosh(\beta v))$
and are thus left with only one parameter, namely $\beta v$. 
The non-bonding component, corresponding to the single particle pole 
$\varepsilon_0$ has a 
finite spectral weight for any $T \neq 0$, while for $T = 0$  it vanishes.

Let us now generalize the above atomic Green's function to the case of a 
lattice. The corresponding Green's function 
$G({\bf k}, i\omega_n)$ 
is obtained by adding \`a la Hubbard the kinetic term which is diagonal 
in ${\bf k}$-space:

\begin{equation}
G({\bf k}, i\omega_n) \; = \; \frac{1}{[ G_{Loc}(i\omega_n) ]^{-1} \; - \;
t_{\bf k}}
\end{equation}
where $t_{\bf k}$ is the Fourier's transformation
of the hopping term\cite{Hubbard}.
The three-pole structure is then repercuted from the atomic Green's function 
onto the lattice Green's function whose dispersion branches are illustrated 
in Fig.(1) and which resemble those previously derived by us on the basis of 
a selfconsistent diagramatic approach\cite{spectral}. 
Using a free parabolic density of states of the form
\begin{equation}
\rho_{0}(\varepsilon) \; = \; \frac{8}{\pi D} \; \sqrt{(\varepsilon/D + 1/2) (
\varepsilon/D - 1/2)}
\; \; \; \mbox{,} \; \; \; -D/2 \leq \varepsilon \leq D/2
\end{equation}
where D denoted the bandwidth of the Fermions $D = 2zt$, 
we obtain  a one-particle Green's function, 
which has the following two-pole structure:

\begin{equation}
G({\bf k}, i\omega_n) \; = \; \frac{1}{2} \left( 1 + \frac{t_{\bf k}}{4 \;
\gamma_{\bf k}} \right) \; \frac{1}{i\omega_n - t_{\bf k} / 2 - \gamma_{\bf k}}
\; + \; \frac{1}{2} \left( 1 - \frac{t_{\bf k}}{4 \; \gamma_{\bf k}} \right)
\; \frac{1}{i\omega_n - t_{\bf k} / 2 + \gamma_{\bf k}}
\end{equation}
 provided the temperature is exactly equal to zero. 
With $\gamma_{\bf k} = \sqrt{t_{\bf k}^{2} / 4 + v^{2}}$, 
this form is reminiscent of the one-particle Green's function exhibiting a 
Bogoliubov like spectrum as in the case of a BCS superconductor. 
The corresponding density of states
 
%insulator
%hybridization spectrum was found in  the previous numerical solutions of
%the Boson-Fermion Model.
%In this approximation, the pseudo-gap in the density of states
%of the fermions corresponds to the formation of a local bounding
%state of two fermions of opposite spins with a boson.

\begin{equation}
\rho(\varepsilon) = \rho_0(\varepsilon - v^2/\varepsilon) 
\left[ \theta(\varepsilon - E_1^-)
\theta( E_1^+ - \varepsilon) + \theta(\varepsilon - E_2^-)
\theta( E_2^+ - \varepsilon) \right]
\end{equation}
with $E_1^{\pm} = \pm \frac{D}{4} 
                + \frac{1}{2}\sqrt{\left( \frac{D}{2} \right)^2 + 4v^2}$
and $E_2^{\pm} = \pm \frac{D}{4} - \frac{1}{2}\sqrt{\left( \frac{D}{2} 
\right)^2 + 4v^2}$
thus shows a gap of size $E_G = 2E_1^-$ at the Fermi level $\varepsilon =0$ 
for $T = 0$. 
Our previous study of the Boson Fermion model clearly showed indications 
for such a gap in form of a dip in the density of states which deepened 
upon lowering the temperature \cite{spectral}.

For any finite temperature $T \neq 0$ the density of states 
$\rho(\varepsilon)$ at the Fermi level turns out to be finite and moreover 
we find
$\rho(0) = \rho_0(0)$. What changes with temperature is the spectral weight 
$Z_f$ of the normal contribution of the single particle excitations. 
It decreases to zero as the temperature decreases which is manifest in a 
gradual narrowing of the peak in $\rho(\varepsilon)$ corresponding to the 
non-bonding one-particle 
state, centered arround the Fermi level
as illustrated in Fig.(2). 
This behaviour can be derived from the expression for the one-particle 
Green's function which in the limit $| \varepsilon | \ll v$
can be derived from the correspondingly approximated spectral function

\begin{equation}
A({\bf k},\varepsilon) = -\frac{1}{\pi}Im G({\bf k},\varepsilon+i0) 
\simeq -\frac{1}{\pi}Im {{Z_f}\over{\varepsilon + i0 - Z_f t_{\bf k}}}
\end{equation}
the density of states is then obtained as
\begin{equation}
\rho(\varepsilon) = \sum_{{\bf k}} A({\bf k},\varepsilon) =
Z_f \int_{-D/2}^{+D/2} d\omega \rho_0(\omega) \delta(\varepsilon - 
Z_f \omega) = \rho_0(\varepsilon/Z_f)
\end{equation}
from which we can see the narrowing down of the peak associated in the 
non-bonding component of the spectrum which scales the free density of states 
$\rho_0$ with the temperature dependent factor $Z_f$.

>From the present discussion we see that the opening of the pseudogap 
obtained in our previous selfconsistent treatment of the Boson-Fermion 
model\cite{spectral} is clearly associated with the disappearence of 
the non-bonding component of the spectrum. As the temperature is decreased 
its spectral weight is transfered into the contributions coming from the 
hybridized Boson-Fermion pairs in bonding and anti-bonding states. 
These contributions to the density of states are centered around 
$\varepsilon_{\pm}$  
(for which moreover we have $max \rho(\varepsilon) = \rho_0(0)$
for $\varepsilon = \varepsilon_{\pm}$, as can be seen directly from Eq.(10)).
The temperature variation of the partial density of states, coming from those
non-bonding contributions, is illustrated in Fig.(3). 
We notice that it scales with $Z_f$ and that it shows a marked saturation 
above a certain characteristic temperature which we identify with the 
temperature  $T^*$ where a pseudogap opens up in the density of states and 
which was previously obtained by us 
using a fully selfconsistent treatment of the Boson-Fermion 
model\cite{spectral,thermo}. 
The saturation above  $T^*$, which is of the order of the coupling constant 
$v$, corresponds to a blocking of the Boson - Fermion-pair exchange mechanism.

The present treatment of the Boson-Fermion model captures the underlying 
physics of the opening of a pseudogap in the density of states due to strong 
local two-particle bonding and anti-bonding resonances which 
increase in spectral weight as the temperature decreases. 
This can be seen from the superfluid pair susceptibility for the Bosons
$\chi_B = - B(i\omega_n =0)$ and for the Fermions
$\chi_F = - G_{FF}(i\omega_n =0)$, which tends to a finite value as $T$ 
decreases to zero. Given the onsite Green's fuctions
$B(\tau) = -\langle T [b_j(\tau) b^{\dagger}_j] \rangle$  and
$G_{FF}(\tau) = -\langle T[c_{j,\downarrow}(\tau) 
c_{j,\uparrow}(\tau)c_{j,\uparrow}^{\dagger}
c_{j,\downarrow}^{\dagger}] \rangle$, the superfluid pair susceptibilities 
for the Bosons and Fermions are determided by
\begin{equation}
\chi_B = \frac{1}{v}{{\beta v + sinh(\beta v)} \over 
{3 + cosh(\beta v)}} \;\; , \;\;
\chi_{F} =\frac{1}{v}{{sinh(\beta v)} \over {3 + cosh(\beta v)}}.
\end{equation}
In the limit $\beta v \Rightarrow \infty$ we hence obtain
$\chi_B \Rightarrow\frac{1}{v}$ and $\chi_F \Rightarrow \frac{1}{v}$,
instead of $\frac{1}{T}$ as for the uncoupled system, which indicates 
the strong two-particle resonance behaviour without any superfluid coherence.
%In the limit where the coupling constant tends to zero we obtain
%$\chi_B \rightarrow \frac{1}{2 T}$ and $\chi_FF \rightarrow \frac{1}{4 T}$
%respectively which indicates the tendency towards a global phase coherent 
%superfluid state, setting in at $T=0$.
%Such features have previously been observed by us on the basis of a fully 
%selfconsistent diagramatic approach to the Boson-Fermion model\cite{spectral}.
A further indication for strong pairing in the normal state comes 
from the expectation value of the number of  pairs
\begin{equation}
n_p = \langle c_{j,\uparrow}^{\dagger} c_{j,\downarrow}^{\dagger} 
c_{j,\downarrow}
c_{j,\uparrow} \rangle = {{1 + cosh(\beta v)} \over {6 + 2 cosh(\beta v)}}
\end{equation}
In the limit of the temperature going to zero, i.e., 
$\beta v \Rightarrow \infty$ we obtain $n_p \Rightarrow \frac{1}{2}$, 
thus showing a significant enhancement over its free particle value, which is 
$\frac{1}{4}$. 

These results, obtained in the atomic limit, clearly demonstrate 
that the pseudogap in the DOS of the Fermions is 
linked to the strong local pair correlations in the normal state with any 
global phase coherence being absent.
 A more refined treatment of those local resonant states  
(such as given in our 
previous study\cite{spectral}  which includes spatial correlations of 
the two-particle resonant states) leads in the normal state to a "PAIR LIQUID"
with fluctuations of a condensate 
of those two-particle states, which ultimately changes the nature of 
the gap leading to a superconducting ground state. 

\section{ACKNOWLEDGEMENT} T.D. would like to acknowledge a grant from 
the French ministry of High Education and Scientific Research and support 
from the Polish Committee of Scientific Research under the project No.
2P03B03111.

$\dagger$ On leave of absence from the Institute of Physics, 
Maria Curie Sklodowska University, Lublin, Poland.

\newpage

%\captions

\begin{figure}
{\bf Fig.1} The dispersion of the three poles of the one-particle Green's 
function in the atomic limitfor $v=0.2$.
\end{figure}

\begin{figure}
{\bf Fig.2} The density of states $\rho(\varepsilon)$ for 
four different temperatures
$T= 0.1$ Fig.(2a), $T= 0.05$ Fig.(2b), $T= 0.001$ Fig.(2c) and  $T= 0$ 
Fig.(2d) which clearly shows the disappearence of the non-bonding component 
as the temperature is decreased.
\end{figure}

\begin{figure}
{\bf Fig.3} Temperature variation of the partial integrated density $I(T)$
 of states arising from the contribution of the non-bonding states.

\end{figure}

\normalsize
\vspace{1cm}
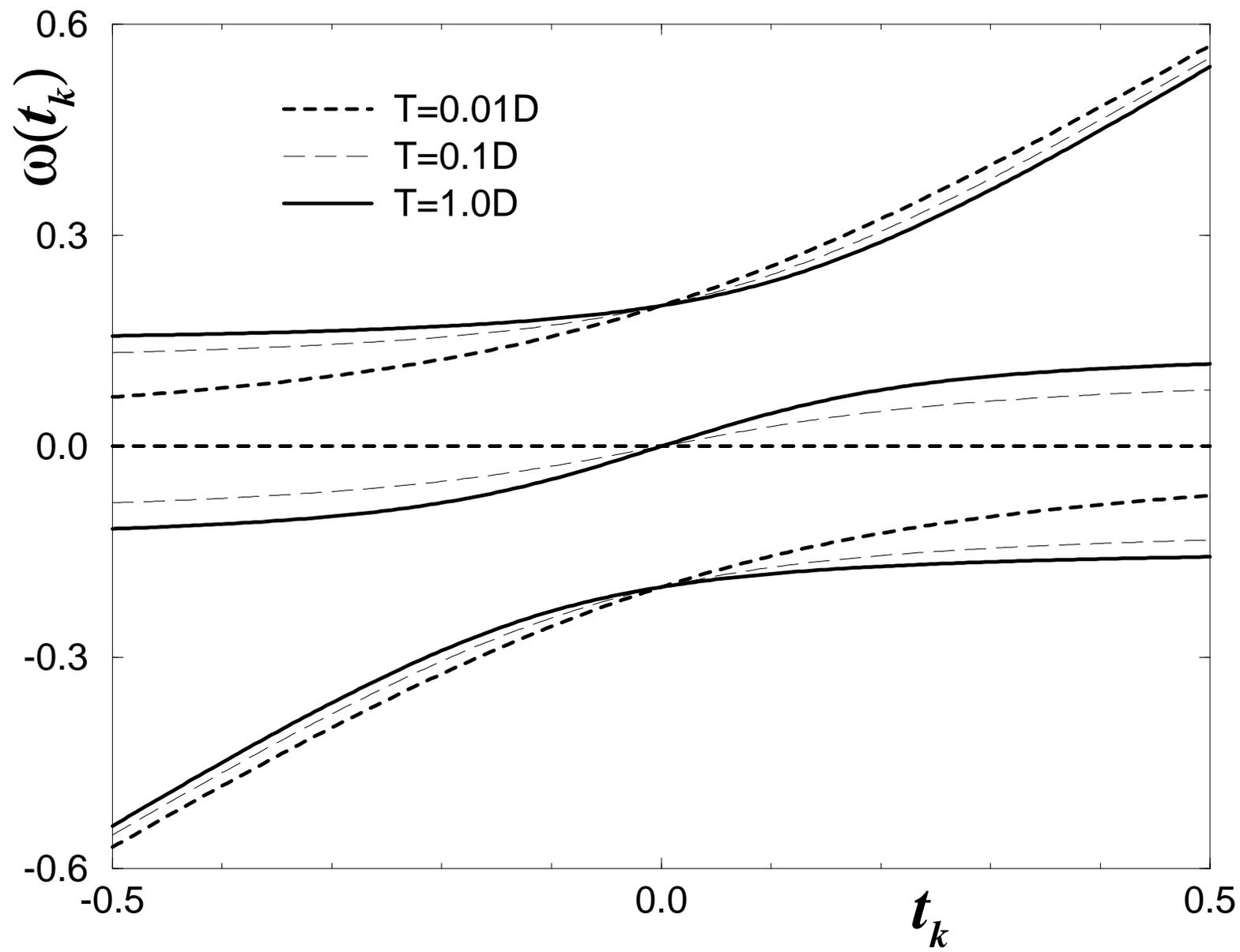
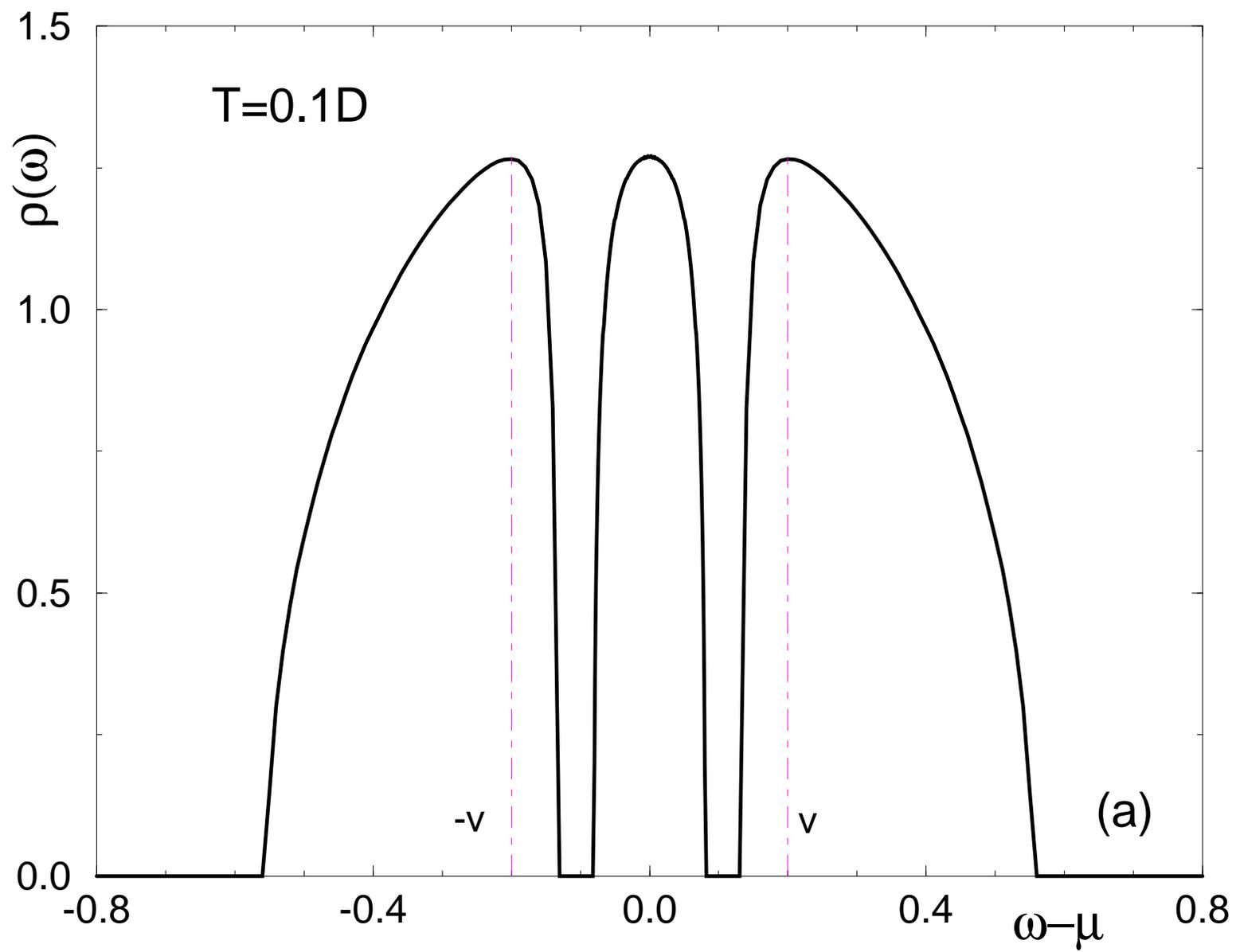
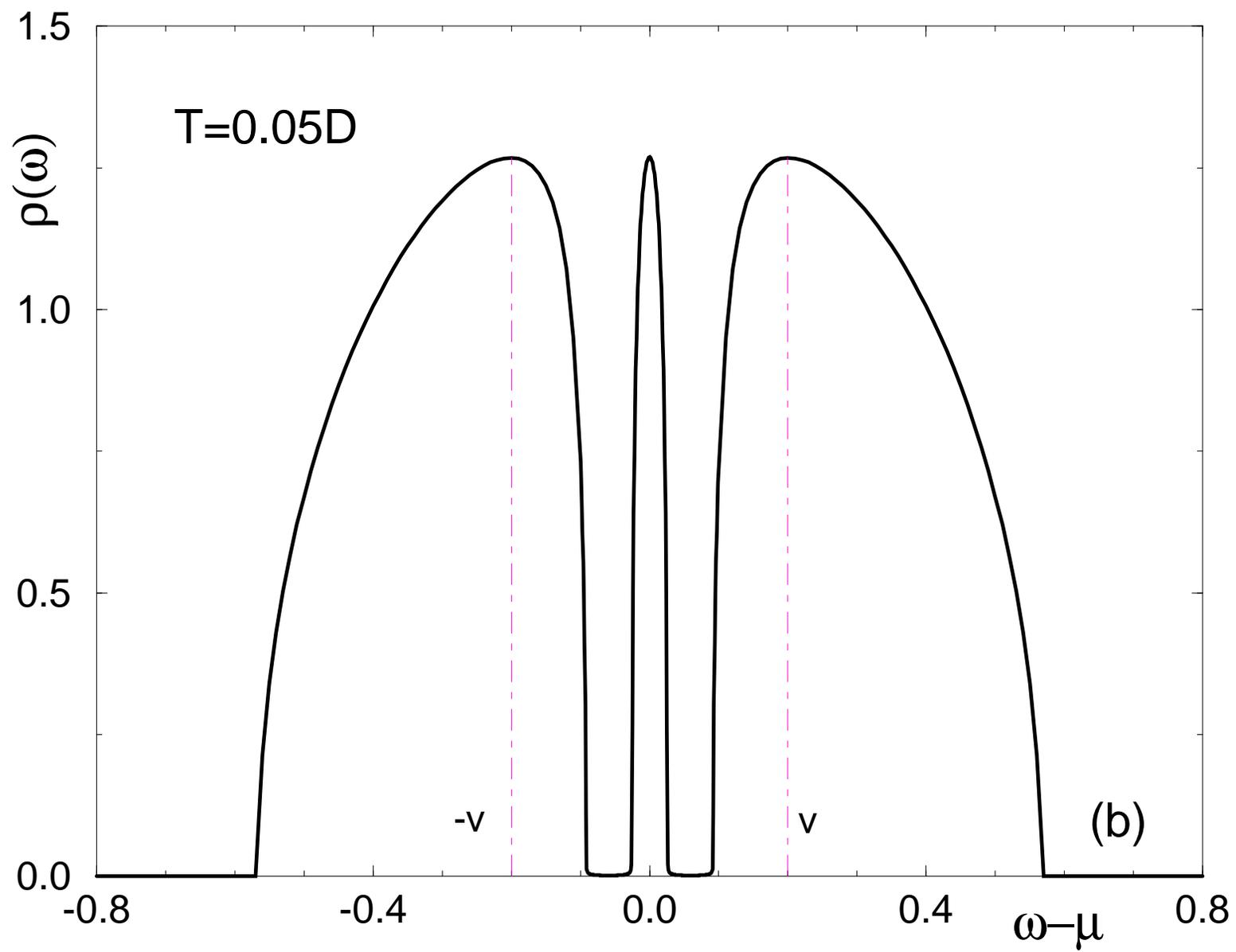
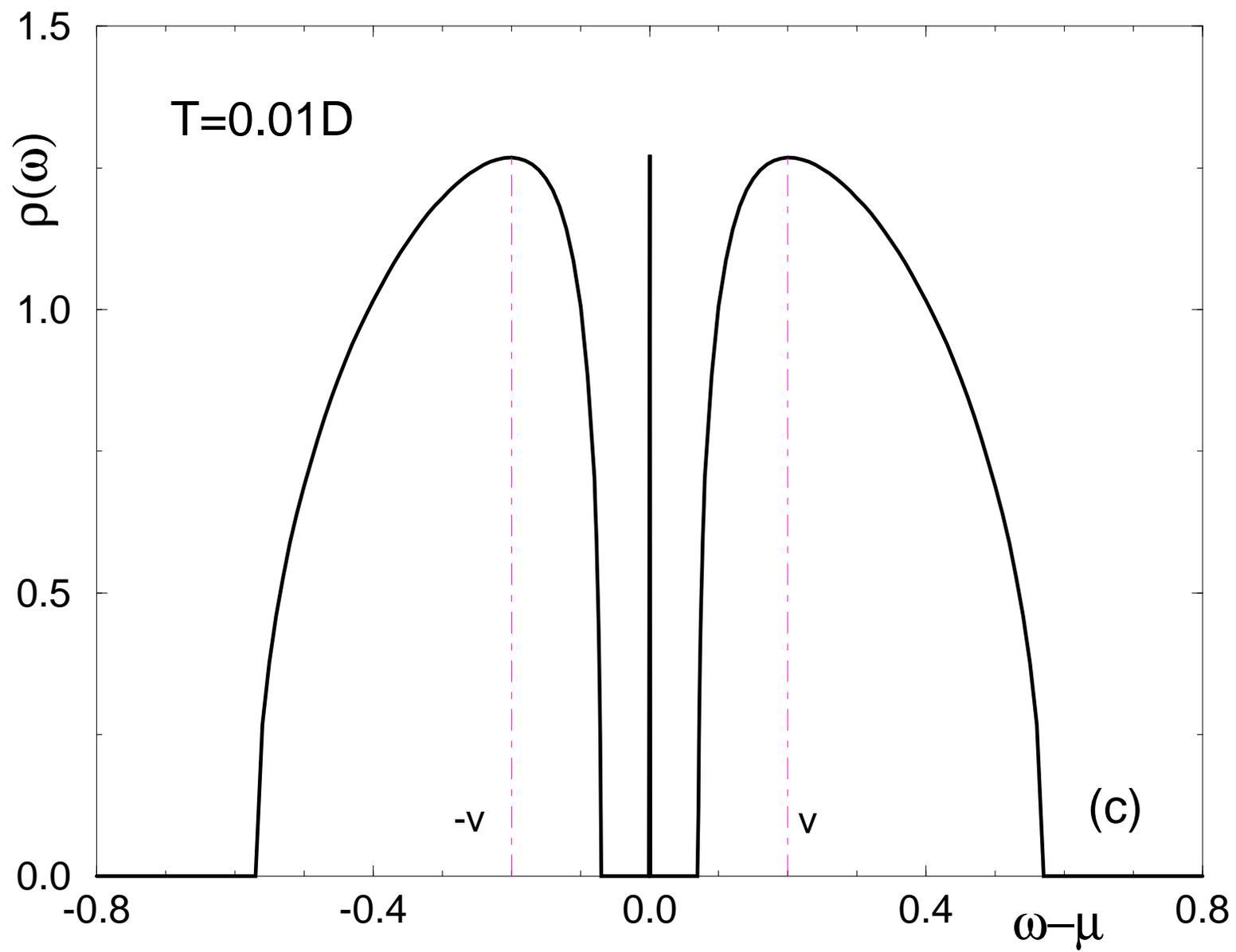
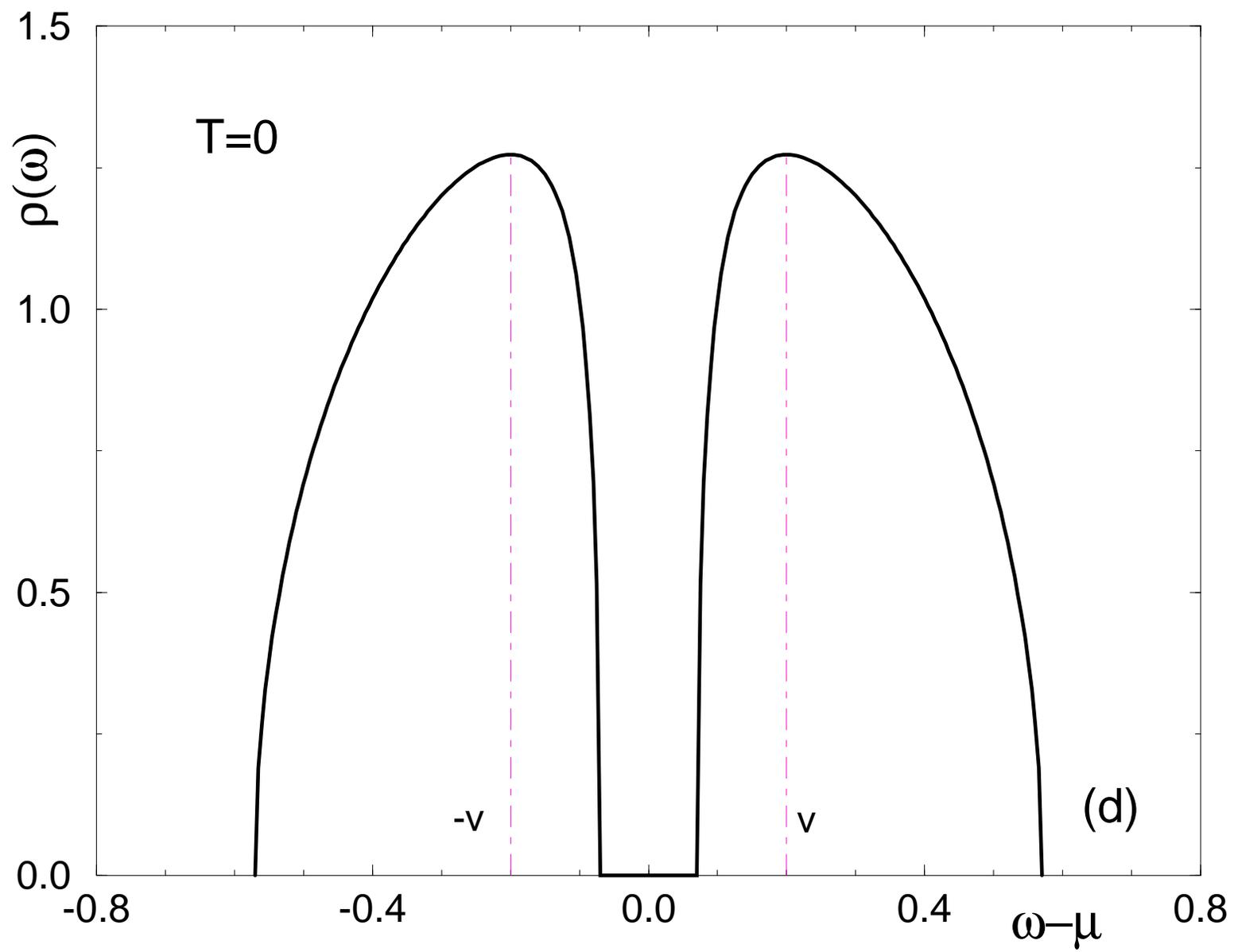
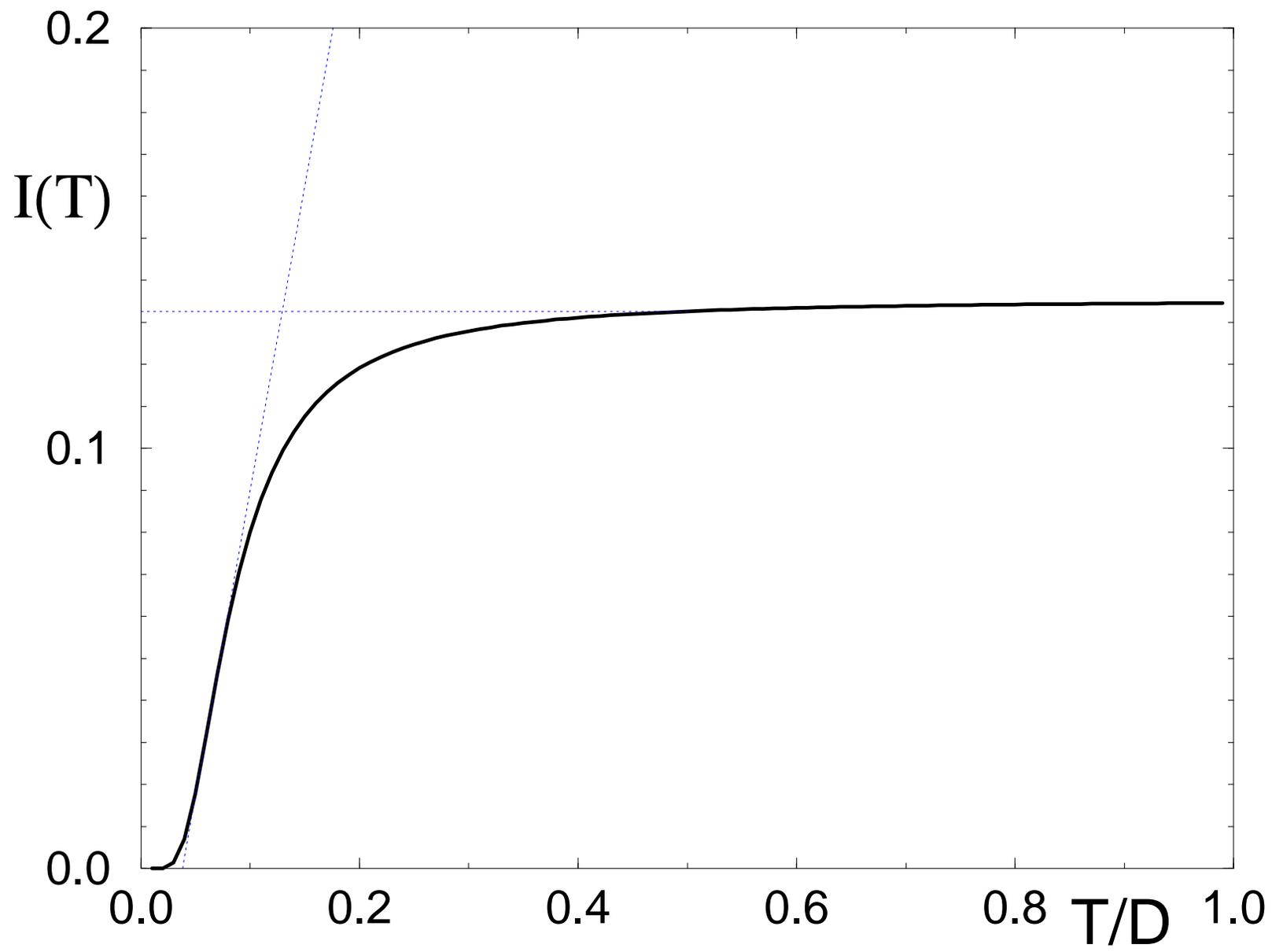
\end{document}